\documentclass[a4paper,12pt]{article}

\usepackage[T1]{fontenc}
\usepackage[utf8]{inputenc}

\usepackage{amsmath}
\usepackage{amssymb}
\usepackage{amsthm}
\usepackage{bm}
\usepackage{color}
\usepackage[margin=2cm]{geometry}
\usepackage{graphicx}
\usepackage{hyperref}
\usepackage{mathrsfs}
\usepackage{subcaption}
\usepackage{natbib}
\usepackage{lmodern}
\usepackage{setspace}
\usepackage{gensymb}
\usepackage{float}
\usepackage{xcolor}
\usepackage{authblk}

\setcitestyle{round}
\title{From Many to One: Consensus Inference in a MIP}

\author[1]{Noel Cressie}
\author[1]{Michael Bertolacci}
\author[1]{Andrew Zammit-Mangion}
\affil[1]{
  School of Mathematics and Applied Statistics

  \vspace{-0.4cm} University of Wollongong, Australia
}

\date{arXiv Version: 8 Jul 21}

\newcommand{\Yvec}{\mathbf{Y}}

\newcommand{\Sigmamat}{\bm{\mathrm{\Sigma}}}

\newcommand{\Xmat}{\bm{\mathrm{X}}}

\newcommand{\Zmat}{\bm{\mathrm{Z}}}

\newcommand{\alphavec}{\bm{\alpha}}

\newcommand{\muvec}{\bm{\mu}}
\newcommand{\etavec}{\bm{\eta}}
\newcommand{\rhovec}{\bm{\rho}}
\newcommand{\sigmavec}{\bm{\sigma}}
\newcommand{\tauvec}{\bm{\tau}}

\newcommand{\bvec}{\mathbf{b}}

\newcommand{\Gau}{\mathrm{Gau}}
\newcommand{\Dist}{\mathrm{Dist}}
\newcommand{\Tmp}{\mathrm{Tmp}}
\newcommand{\var}{\mathrm{var}}
\newcommand{\cov}{\mathrm{cov}}
\newcommand{\MSPE}{\mathrm{MSPE}}

\newcommand{\nospellcheck}[1]{#1}

\definecolor{darkgreen}{RGB}{0, 120, 0}

\begin{document}

\maketitle

\begin{abstract}
  A Model Intercomparison Project (MIP) consists of teams who each estimate the same underlying quantity (e.g., temperature projections to the year 2070), and the spread of the estimates indicates their uncertainty. It recognizes that a community of scientists will not agree completely but that there is value in looking for a consensus and information in the range of disagreement. A simple average of the teams' outputs gives a consensus estimate, but it does not recognize that some outputs are more variable than others. Statistical analysis of variance (ANOVA) models offer a way to obtain a weighted consensus estimate of outputs with a variance that is the smallest possible and hence the tightest possible `one-sigma' and `two-sigma' intervals. Modulo dependence between MIP outputs, the ANOVA approach weights a team's output inversely proportional to its variation. When external verification data are available for evaluating the fidelity of each MIP output, ANOVA weights can also provide a prior distribution for Bayesian Model Averaging to yield a consensus estimate. We use a MIP of carbon dioxide flux inversions to illustrate the ANOVA-based weighting and subsequent consensus inferences.
\end{abstract}


\section{Introduction}
\label{sec:introduction}

Increasing our knowledge of the behavior of geophysical processes in the past, at the present, and in the future requires a scientific understanding of the processes, of the parameters that govern them and of the observing systems measuring them. That understanding can be quantified through a geophysical model and the mathematical and statistical relationships embedded in it. The presence of uncertainty in the model formulation is generally recognized, although where and how it enters is usually a topic of more or less disagreement \citep[e.g.,][]{knuttietal2010}. Reconciliation and consensus estimation in these settings is a topic of central interest in many geophysical studies of global import, including the Coupled Model Intercomparison Projects \citep[CMIP,][]{meehletal2000}, and the Ice Sheet Mass Balance Inter-comparison Exercises \citep[IMBIE,][]{shepherdetal2018}.

This article considers the Model Intercomparison Project (MIP) approach, commonly used to understand the similarities and differences in competing geophysical models. Here a number of teams produce outputs that estimate the underlying geophysical process in question, but the outputs generally differ due to modeling choices or assumptions. The collection of outputs, which we simply call the `MIP's outputs,' is sometimes also called a `multi-model ensemble'. When analyzing and summarizing a MIP's outputs, it is common to use the ensemble average \citep[`one vote per model'; see][Section 9.2.2.3]{flatoetal2014}, and either the range from the ensemble's lowermost to its uppermost or the ensemble standard deviation, to put bounds on how the process did, does, or will behave. However, it is recognized that this practice of treating all outputs equally is frequently inappropriate \citep{tebaldiknutti2007}, as some outputs reproduce observational data more faithfully than others and should therefore receive more attention \citep{knutti2010}. At the same time, similarities between the methodologies of certain teams suggest that these team's outputs should not be treated as completely independent and therefore each one should receive less attention \citep{tebaldiknutti2007,abramowitzetal2019}. The most common paradigm (and the one used in this article) for converting these ideas into a method for combining outputs is to \emph{weight the outputs} according to given criteria. Under this paradigm, the ensemble is typically summarized through a weighted mean rather than the unweighted mean (i.e., the average), and the uncertainty quantification also takes into account the weights \citep[e.g.,][]{tebaldietal2005}.

\begin{figure}
  \begin{center}
    \includegraphics{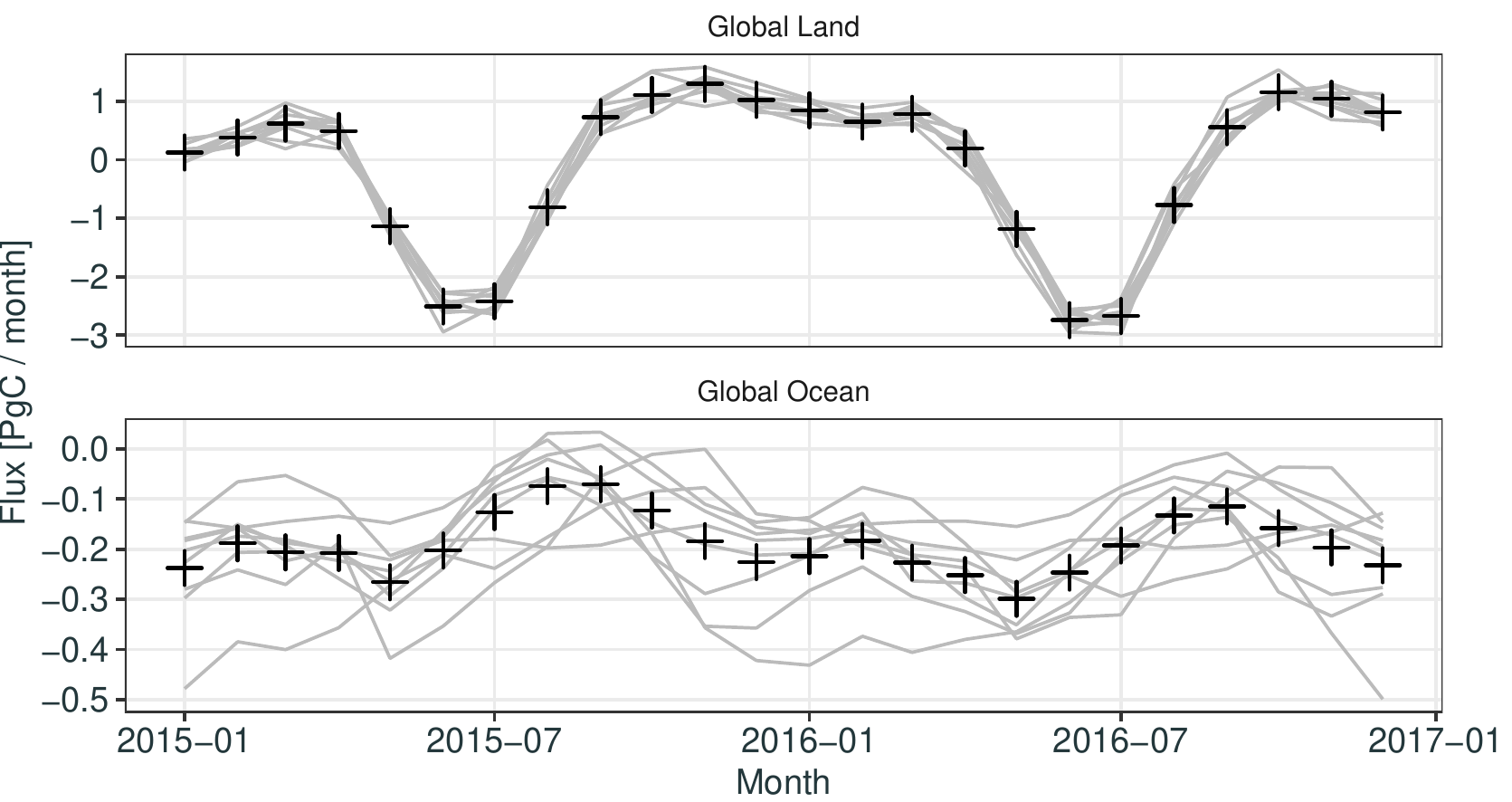}
  \end{center}

  \caption{
    Global land (top) and global ocean (bottom) estimated CO$_2$ fluxes from the nine teams who participated in the OCO-2 Model Intercomparison Project (MIP), reported in \citet{crowelletal2019}. The lines show the fluxes of each ensemble member, and the crosses show the unweighted ensemble monthly means from January 2015 through December 2016. The fluxes are for the LN observation type (see Section~\ref{sec:oco2_mip}).
  }
  \label{fig:flux_land_ocean_ln}
\end{figure}

A key question is how to construct the ensemble weights. The methods used to do so depend on the details of the application. It is natural to use any available observational data to help with the weighting and, broadly speaking, applications fall onto a spectrum depending on how much verification data are available. At one end of the spectrum, when the underlying geophysical process is contemporaneous, such as in the prediction of meteorological fields, weighting methods can typically take advantage of large amounts of verification data \citep[e.g.,][]{krishnamurtietal2000}. At the other end of the spectrum, when the geophysical process is extremely difficult to measure directly, little-to-no direct verification data may be available; this situation is the focus of this article.

In the middle of the spectrum, limited verification data are available. The archetypal case of this is the projection of future climate, where verification data are available for a number of years up to the beginning of the projections. For example, see Figure~11.9 on page 981 of the Intergovernmental Panel on Climate Change Fifth Assessment Report \citep{kirtmanetal2013}, which includes 42 projections from the CMIP5 ensemble members. For 1986--2005, the projections are based on historical observed greenhouse gas concentrations. Afterwards, they are based on a pre-specified path of future emissions, called a representative concentration pathway (RCP, in this case RCP 4.5). Verification data are available for the years 1986--2005, and they are used to evaluate the performance of the models. In this and similar settings, there are many reasons why historical performance may not guarantee future accuracy \citep{reifentoumi2009} but, nonetheless, one should not discount this historical information entirely \citep{knutti2010}.

For MIPs like CMIP5, where limited verification data are available, \citet{giorgimearns2002} present a framework called the Reliability Ensemble Average (REA), in which heuristic weights are derived that reward ensemble members that are simultaneously (i) in consensus and (ii) in agreement with verification data. \citet{tebaldietal2005} formalize REA in a statistical framework for analyzing model projections of future global temperatures across 22 regions. In their approach, the regions are treated independently, and the individual outputs are modeled as random quantities that center on the true climate state. The variances differ between the outputs, which imply different weights, and the estimated weights balance the REA criteria (i) and (ii). There are several extensions of the \citet{tebaldietal2005} method: \citet{smithetal2009} treat the regions jointly and describe a cross-validation technique to validate the model fit, while \citet{tebaldisanso2009} allow for joint modeling of both temperature and precipitation.

A key limitation of the REA framework and its associated methods is that they do not account for dependence between outputs \citep{abramowitzetal2019}. More recent developments have attempted to address this by modifying the weights, typically by down-weighting sub-groups of outputs that are considered to be dependent \citep{bishopabramowitz2013}. However, there is disagreement on what it means for certain MIP outputs to be dependent, and approaches to understanding this include the distance between outputs \citep[e.g.,][]{sandersonetal2017}, the historical `genealogy' of climate models \citep[e.g.,][]{knuttietal2013,knuttietal2017}, component-wise similarity between models \citep[e.g.,][]{boe2018}, dependence between errors \citep[e.g.,][]{bishopabramowitz2013,haughtonetal2015}, and probabilistic/statistical dependence \citep[e.g.,][]{chandler2013,sunyeretal2014}.

We now consider the case that is our main focus, where little to no \emph{direct} verification data are available. An example of this, described in detail in Section~\ref{sec:oco2_mip}, is the estimation of surface sources and sinks (fluxes) of CO$_2$ in a MIP; see Figure~\ref{fig:flux_land_ocean_ln}, which shows the nine teams' estimates of global land and global ocean non-fossil-fuel fluxes, as well as the unweighted ensemble average. Fluxes of CO$_2$ occur over large temporal and geographical scales and are difficult to measure directly. One way their distribution is inferred indirectly is through their consequences on atmospheric concentrations of CO$_2$, for which observations are available \citep{ciaisetal2010}. However, there are complications in using such indirect data taken anywhere in the atmosphere for verification, because issues such as overfitting and model misspecification mean that an accurate reproduction of the observed concentrations may not imply accurate fluxes \citep[e.g.,][]{houwelingetal2010}. Therefore, in this case and in cases like it, one is not applying \nospellcheck{REA's} criterion (ii), of agreement with flux data. However, according to REA criterion (i), one can infer the consensus between the MIP's outputs (while taking into consideration model dependence, if needed). To apply the first criterion, we present a framework that enables \emph{statistically unbiased prediction and estimation} (SUPE) based on an \emph{analysis of variance} (ANOVA) statistical model, originally used for analyzing agricultural field experiments \citep{fisher1935,snedecor1937}. The framework promotes models that are in agreement and discounts those that are in disagreement, but it does so selectively in regions and seasons thought to have their own geophysical characteristics. For each of these regions/seasons, it provides a consensus summary and accompanying uncertainty quantification. It can also cater for model interdependencies when generating the summary, provided these have been, or can be, identified from \emph{a priori} or expert knowledge.

Section~\ref{sec:statistical_anova} presents the ANOVA model in general, by recognizing that different factor combinations (e.g., spatial regions, seasons, geophysical-model configurations) will lead to agreement in differing degrees. We model the teams' outputs for each factor combination as having different variances and, in Section~\ref{sec:optimal_weighting}, obtain statistically optimal weights based on these variances that allow optimal SUPE and its uncertainty quantification. In Section~\ref{sec:estimation}, we show how to estimate these weights using likelihood methods derived from the statistical distributions in the ANOVA model. Section~\ref{sec:oco2_mip} gives an illustration of our statistical-weighting methodology on the MIP published by \citet{crowelletal2019}, where nine teams produced outputs of global carbon-dioxide fluxes that were compared and summarized. A discussion of our framework, which we call SUPE-ANOVA, and conclusions, are given in Section~\ref{sec:discussion}.

\section{Analysis of variance (ANOVA) for a MIP analysis}
\label{sec:statistical_anova}

The area of statistics known as Experimental Design was born out of a need to control variations in agricultural field experiments, in order to compare and then select crop treatments based on replicate observations. The statistical ANOVA model was devised to analyze these experiments, and much has been written about it, starting with \citet{fisher1935}, \citet{snedecor1937}, \nospellcheck{Snedecor} and Cochran in its eight editions (1957, \dots, \citeyear{snedecor2014}), and many other texts. It has been adopted in a number of disciplines driven by experimentation, including industrial manufacturing \citep[e.g.,][]{wuhamada2000}, and computer experiments where the so-called treatments are complex algorithms and code \citep[e.g.,][]{santnereal2010}. We propose using the ANOVA framework for the analysis of outputs from important \emph{geophysical experiments} performed within a MIP. It enables the output from \emph{many} to be weighted in a statistically efficient manner to obtain \emph{one} consensus estimate and a quantification of its uncertainty.

Let `$f$' represent a combination of \emph{factors} (e.g., season, region) that the underlying geophysical process is expected to depend on, and let `$j$' index the output of the various teams participating in the MIP. By stratifying on each factor combination, we can write the output for the $f$th factor combination ($f = 1, \ldots, F$) and the $j$th team ($j = 1, \ldots, J$) as
\begin{equation}
  \left\{ Y_{f, i}^{(j)} : i = 1, \ldots, I(f) \right\}, \quad \text{for } f = 1, \ldots, F, \enskip j = 1, \ldots, J,
  \label{eqn:mip_outputs}
\end{equation}
where $I(f)$ represents the number of replicates for factor combination $f$. Each combination $(f, i)$ corresponds to a particular region/time-period in the MIP, and note that $I(f)$ may differ for different factor combinations, reflecting the ability to capture complex MIP designs. In the analysis of the CO$_2$-flux MIP given in Section~\ref{sec:oco2_mip}, $I(f)$ is a constant number (equal to six) for all regions and seasons.

The notation recognizes that each output is split into $F$ blocks (factor combinations) and, within the $f$th block, there are $(I(f) \cdot J)$ outputs that are observing the same part of the geophysical process, corresponding to the $I(f)$ replicates and the $J$ teams participating in the MIP. For example, suppose the MIP's goal is to compare $20$ climate models' projections of North American mean temperature between the years $2021$ and $2070$, inclusive, at yearly intervals. Say the factor $f$ corresponds to decade, so $f \in \{ 1, 2, \ldots, 5 \}$. Further, $i \in \{ 1, 2, \ldots, 10 \}$ corresponds to the $I(f) = 10$ years within a decade, and $j \in \{ 1, 2, \ldots, 20 \}$ corresponds to the $20$ climate models. For climate model 17 (say) and decade 4 (say, which corresponds to the years between $2051$ to $2060$, inclusive), the output at yearly intervals within the decade is $\{ Y_{4, i}^{(17)} : i = 1, \ldots, 10 \}$.

The statistical ANOVA model proposed for \eqref{eqn:mip_outputs} is, for $f = 1, \ldots, F$, $i = 1, \ldots, I(f)$, and $j = 1, \ldots, J$,
\begin{equation}
  Y_{f, i}^{(j)} = Y_{f, i} + \eta_{f, i}^{(j)},
  \label{eqn:anova_team}
\end{equation}
where $\{ \eta_{f, i}^{(1)}, \ldots, \eta_{f, i}^{(J)} \}$ each have mean zero, $\var(\eta_{f, i}^{(j)}) = (\sigma_f^{(j)})^2$, and $\cov(\eta_{f, i}^{(j)}, \eta_{f, i}^{(j')}) = \rho_f^{(j, j')} \sigma_f^{(j)} \sigma_f^{(j')}$. Here, $Y_{f, i}$ is the MIP's consensus of the geophysical process that each of the MIP outputs is attempting to estimate, and $\eta_{f, i}^{(j)}$ is the deviation from the consensus of the $j$th output, assumed to have team-specific variances and covariances and to be independent of $Y_{f, i}$. Marginally, $\eta_{f, i}^{(j)} \sim \Dist( 0, (\sigma_f^{(j)})^2)$, where $\Dist(\mu, \sigma^2)$ denotes a generic distribution with mean $\mu$ and variance $\sigma^2$. Dependence between teams' outputs is captured with correlations $\{ \rho_f^{(j, j')} : j, j' = 1, \ldots, J \}$. The statistical ANOVA model has variances that may differ from team to team. A mathematical-statistical result called the Gauss--Markov theorem \citep[e.g.,][]{casellaberger2006} implies that the hereto-referred weights are inversely proportional to these variances, and there is a more general version of the theorem that includes the dependencies between outputs; see Appendix~\ref{sec:general_equations}.

For $f = 1, \ldots, F$ and $i = 1, \ldots, I(f)$, we write the consensus as,
\begin{equation}
  Y_{f, i} = \mu_f + \alpha_{f, i},
  \label{eqn:anova_underlying}
\end{equation}
where $\{ \alpha_{f, 1}, \ldots, \alpha_{f, I(f)} \}$ each have mean zero. Here, $\mu_f$ is the unknown mean of the $I(f)$ replicates, which we refer to as the climatological mean, and $\{ \alpha_{f, 1}, \ldots, \alpha_{f, I(f)} \}$ are random effects that capture variability in the replicates, and whose distribution we denote as $\Dist(0, \tau^2_f)$. In the illustrative example of North American temperatures, $f$ indexes decade, $\mu_f$ corresponds to the consensus temperature in the $f$th decade, while $\alpha_{f, i}$ is the deviation from $\mu_f$ for the $i$th year within the decade. The variability of temperatures can differ between decades, reflected by $\tau^2_f$ for $f = 1, \ldots, 5$. A temporal trend within the $f$th decade could also be included in the model for $\{ \alpha_{f, i} : i = 1, \ldots, I(f) \}$. The team-specific variances $\{ (\sigma_f^{(j)})^2 : j = 1, \ldots, J \}$ and correlations $\{ \rho_f^{(j, j')} : j, j' = 1, \ldots, J \}$ defined through \eqref{eqn:anova_team} may also be different for different decades. In the North American temperature example, a team's output with higher weight in one decade may not have higher weight in another.

The statistical ANOVA model in \eqref{eqn:anova_team} and \eqref{eqn:anova_underlying} allows statistically unbiased prediction and estimation (SUPE) on the consensus for the climatological means, $\{ \mu_f : f = 1, \ldots, F \}$, and on the consensus for the underlying geophysical process, $\{ Y_{f, i} : i = 1, \ldots, I(f); \enskip f = 1, \ldots, F \}$. Notice the heterogeneity of variation in \eqref{eqn:anova_team} and \eqref{eqn:anova_underlying}, where the variances (and correlations) depend on both $f$ and $j$. In the next section, we explain how knowledge of these variance (and correlation) parameters leads to statistically optimal weighting of the $J$ MIP outputs \eqref{eqn:mip_outputs} for each $f$ when making inference on $\{ \mu_f \}$ and $\{ Y_{f, i} \}$.

\section{Weighting MIP outputs for optimal inference}
\label{sec:optimal_weighting}

In the simple example from Section~\ref{sec:statistical_anova}, of combining 20 climate models, not all models are expected to show the same variability. Different assumptions are made, say, about the model's base resolution at which energy and water exchange takes place, how nonlinear dynamics are approximated, and so forth. This possibility is taken into account through \eqref{eqn:anova_team}, where the outputs from each team receives a different variance. A consequence of this is that, for optimal estimation and prediction of the consensus, different weights should be assigned to the different teams. This section discusses how to construct the optimal weights.

To illustrate optimal weighting in a simpler setting, without factor combinations, consider each climate model's 2070 projected North American mean temperature (\nospellcheck{Tmp}), $Y_\Tmp^{(1)}, \ldots, Y_\Tmp^{(20)}$. In this case, there is only one replication, so we omit the subscript $i$ on all quantities. To start with, assume a \nospellcheck{na\"{i}ve} statistical model wherein the 20 climate models all have the \emph{same} variability:
\begin{equation}
  Y_\Tmp^{(j)} = \mu_\Tmp + \eta_\Tmp^{(j)}, \enskip  j = 1, \ldots, 20,
  \label{eqn:y_j_nat}
\end{equation}
where independently, $\eta_\Tmp^{(j)} \sim \Dist(0, \sigma^2_\Tmp)$. For simplicity of exposition, the correlations in this example are assumed to be zero. From \eqref{eqn:y_j_nat}, $Y_\Tmp^{(j)} \sim \Dist(\mu_\Tmp, \sigma^2_\Tmp)$ for $j = 1, \ldots, 20$. Then the \emph{best linear unbiased estimator (BLUE)} of $\mu_\Tmp$ is the unweighted mean, and its variance is straightforward to calculate. That is,
\begin{equation}
  \bar{\mu}_\Tmp \equiv \frac{1}{20} \sum_{j = 1}^{20} Y_\Tmp^{(j)}, \enskip
  \text{and } \var(\bar{\mu}_\Tmp) = \sigma^2_\Tmp / 20.
  \label{eqn:nat_homo_blue}
\end{equation}
To justify using the word `best' in the acronym `BLUE,' one needs to build a statistical model. Assuming the model \eqref{eqn:y_j_nat}, the estimate \eqref{eqn:nat_homo_blue} is the most precise linear estimator, in that $\var(\bar{\mu}_\Tmp) \leq \var(\tilde{\mu}_\Tmp)$, for any other linear unbiased estimator, $\tilde{\mu}_\Tmp$, of $\mu_\Tmp$. However, \eqref{eqn:y_j_nat} does not allow for different variabilities in the different MIP outputs.

Now, for the sake of argument, suppose that the first 15 out of the 20 climate models have an underlying resolution of $0.5 \degree \times 0.5 \degree$, while the remaining five models have a resolution of $1 \degree \times 1 \degree$. Consequently, an assumption of equal variance in the statistical distributions in \eqref{eqn:y_j_nat} will need modification. Assume variances $(\sigma_\Tmp^{(\text{lo})})^2$ and $(\sigma_\Tmp^{(\text{hi})})^2$ are known, and
\begin{equation}
  Y_\Tmp^{(j)} \sim \begin{cases}
    \Dist\left( \mu_\Tmp, \left( \sigma_\Tmp^{(\text{lo})} \right)^2 \right); & j = 1, \ldots, 15, \\
    \Dist\left( \mu_\Tmp, \left( \sigma_\Tmp^{(\text{hi})} \right)^2 \right); & j = 16, \ldots, 20.
  \end{cases}
  \label{eqn:y_j_nat_hetero}
\end{equation}
In Section~\ref{sec:estimation}, we show how these variances can be estimated. The parameter $\mu_\Tmp$ is still the target of inference but, under the statistical model \eqref{eqn:y_j_nat_hetero}, the Gauss--Markov theorem \citep[e.g.,][]{casellaberger2006} can be used to establish that the estimate $\bar{\mu}_\Tmp$ in \eqref{eqn:nat_homo_blue} is no longer the BLUE. We define the weights for estimating the projected mean temperature in the year 2070, $\mu_\Tmp$, as follows:
\begin{equation}
  w_\Tmp^{(j)} = \begin{cases}
    \left( \sigma_\Tmp^{(\text{lo})} \right)^{-2}
    \left\{ 15 \left( \sigma_\Tmp^{(\text{lo})} \right)^{-2} + 5 \left( \sigma_\Tmp^{(\text{hi})} \right)^{-2} \right\}^{-1}; & j = 1, \ldots, 15, \\
    \left( \sigma_\Tmp^{(\text{hi})} \right)^{-2}
    \left\{ 15 \left( \sigma_\Tmp^{(\text{lo})} \right)^{-2} + 5 \left( \sigma_\Tmp^{(\text{hi})} \right)^{-2} \right\}^{-1}; & j = 16, \ldots, 20.
  \end{cases}
  \label{eqn:mu_tmp_weights}
\end{equation}

Under the statistical model \eqref{eqn:y_j_nat_hetero}, the Gauss--Markov theorem can be used to establish that the BLUE (the most precise linear unbiased estimator) and its variance are, respectively,
\begin{equation}
  \mu_\Tmp^* = \sum_{j = 1}^{20} w_\Tmp^{(j)} Y_\Tmp^{(j)}, \enskip
  \text{and } \var(\mu_\Tmp^*) = \left\{ 15 \left( \sigma_\Tmp^{(\text{lo})} \right)^{-2} + 5 \left( \sigma_\Tmp^{(\text{hi})} \right)^{-2} \right\}^{-1}.
  \label{eqn:mu_tmp_blue}
\end{equation}
That is, the statistically optimal weights for estimation of $\mu_\Tmp$ are given by \eqref{eqn:mu_tmp_weights}, and the best estimator of $\mu_\Tmp$ is obtained by weighting each datum inversely proportional to its variance.

A more general statistical model, analogous to what we consider in \eqref{eqn:anova_underlying}, involves the \emph{random} effect, $\alpha_\Tmp$, as follows:
\begin{equation}
  Y_\Tmp = \mu_\Tmp + \alpha_\Tmp,
\end{equation}
where $\alpha_\Tmp \sim \Dist(0, \tau_\Tmp^2)$, and $\tau_\Tmp^2$ is assumed known. When $\alpha_\Tmp = 0$, we recover the model \eqref{eqn:y_j_nat_hetero}, however the random effect $\alpha_\Tmp$ plays an important role in recognizing the stochasticity in the projected temperature. Hence, for $j = 1, \ldots, 20$,
\begin{equation}
  Y_\Tmp^{(j)} = Y_\Tmp + \eta_\Tmp^{(j)} = \mu_\Tmp + \alpha_\Tmp + \eta_\Tmp^{(j)}.
\end{equation}
Notice that there are between-model correlation terms induced by the random effect; for $j \neq j'$,
\[
  \cov\left( Y_\Tmp^{(j)}, Y_\Tmp^{(j')} \right) = \cov\big( \alpha_\Tmp + \eta_\Tmp^{(j)}, \alpha_\Tmp + \eta_\Tmp^{(j')} \big) = \tau^2_\Tmp,
\]
where recall that in this simple example $\eta_\Tmp^{(j)}$ and $\eta_\Tmp^{(j')}$ are uncorrelated. \emph{Prediction} of $Y_\Tmp$, which is the appropriate inference in the presence of the random effect, is made through the predictive distribution of $Y_\Tmp$ given all the data from all the climate-model outputs; the optimal predictor of $Y_\Tmp$ is the mean of this predictive distribution. We use the \emph{best linear unbiased predictor (BLUP)} to make inference on $Y_\Tmp$. In the simple example of North American temperatures, the BLUP is
\begin{equation}
  Y_\Tmp^* = \lambda_\Tmp^{(0)} \mu_\Tmp + \sum_{j = 1}^{20} \lambda_\Tmp^{(j)} Y_\Tmp^{(j)},
  \label{eqn:Y_tmp_blup}
\end{equation}
where
\begin{equation}
  \lambda_\Tmp^{(j)} = \begin{cases}
    \tau_\Tmp^{-2} \left\{ \tau_\Tmp^{-2} + 15 \left( \sigma_\Tmp^{(\text{lo})} \right)^{-2} + 5 \left( \sigma_\Tmp^{(\text{hi})} \right)^{-2} \right\}^{-1}; & j = 0, \\
    \left( \sigma_\Tmp^{(\text{lo})} \right)^{-2} \left\{ \tau_\Tmp^{-2} + 15 \left( \sigma_\Tmp^{(\text{lo})} \right)^{-2} + 5 \left( \sigma_\Tmp^{(\text{hi})} \right)^{-2} \right\}^{-1}; & j = 1, \ldots, 15, \\
    \left( \sigma_\Tmp^{(\text{hi})} \right)^{-2} \left\{ \tau_\Tmp^{-2} + 15 \left( \sigma_\Tmp^{(\text{lo})} \right)^{-2} + 5 \left( \sigma_\Tmp^{(\text{hi})} \right)^{-2} \right\}^{-1}; & j = 16, \ldots, 20.
  \end{cases}
  \label{eqn:Y_tmp_weights}
\end{equation}
Notice that, for both optimal prediction (BLUP) and optimal estimation (BLUE), outputs with larger variability around the consensus receive lower weights. Additionally, for the BLUP given by \eqref{eqn:Y_tmp_blup}, the weight on $\mu_\Tmp$ depends on how variable the underlying process is; that is, it depends on $\tau_\Tmp^2$. Thus, the more variable the geophysical process is (captured through the parameter $\tau^2_\Tmp$), the less weight its mean $\mu_\Tmp$ receives in the optimal predictor, $Y_\Tmp^*$, of $Y_\Tmp$.

To justify the word `best' in the acronym `BLUP,' an uncertainty measure called the mean-squared prediction error (MSPE) is used. Let $\tilde{Y}_\Tmp$ be any linear predictor of $Y_\Tmp$; then the BLUP, $Y_\Tmp^*$, minimizes the MSPE since
\[
  E(Y_\Tmp^* - Y_\Tmp)^2 \leq E(\tilde{Y}_\Tmp - Y_\Tmp)^2.
\]
A general formula for the MSPE of the BLUP is given later in this section.

Returning to the generic MIP, the steps to carry out optimally weighted linear inference follow the same line of development as in the temperature example: (1) specify the first-moment and second-moment parameters in the ANOVA model; (2) derive the variances and covariances of all random quantities in terms of the parameters; (3) weight the outputs (essentially) inversely proportional to their variances to obtain a BLUE/BLUP (i.e., to obtain a best SUPE); (4) quantify uncertainty (variance of the BLUE and MSPE of the BLUP). These four steps require an initialization that might be called a zeroth step, namely the estimation of the parameters $(\sigma_f^{(1)})^2, \ldots, (\sigma_f^{(J)})^2$ and $\tau^2_f$; see Section~\ref{sec:estimation}. The correlation parameters, $\{ \rho_f^{(j, j')} : j, j' = 1, \ldots, J \}$, could be known \emph{a priori} or estimated using a likelihood-based methodology along with the variance parameters; see Appendix~\ref{sec:general_equations}. All together, these steps make up an inference framework for MIPs that we call SUPE-ANOVA.

We now give the optimal estimation and prediction formulas for the generic statistical ANOVA model under the assumption that the correlations between the models, $\{ \rho_f^{(j, j')} \}$, are zero; the general case of non-zero correlations involves matrix algebra and is given in Appendix~\ref{sec:general_equations}. From \eqref{eqn:anova_team} and \eqref{eqn:anova_underlying}, it is straightforward to show that $E(Y_{f, i}^{(j)}) = \mu_f$, $\var(Y_{f, i}^{(j)}) = \tau^2_f + (\sigma_f^{(j)})^2$, and $\cov(Y_{f, i}^{(j)}, Y_{f, i}^{(j')}) = \tau^2_f$. From a multivariate version of the Gauss-Markov theorem \citep[e.g.,][]{rencherchristensen2012}, the BLUE of $\mu_f$, and its variance, are given by
\begin{align}
  \mu_f^*
  & \equiv \sum_{j = 1}^J \sum_{i = 1}^{I(f)} w_{f, i}^{(j)} Y_{f, i}^{(j)},
  \label{eqn:mu_blue} \\
  S^2_f
  & \equiv
    \var(\mu_f^*)
    =
    \frac{1}{I(f)}\left(
    \frac{1}{\sum_{j = 1}^J \left( \sigma_f^{(j)} \right)^{-2}} + \tau_f^2
  \right),
  \label{eqn:mu_blue_var}
\end{align}
for $f = 1, \ldots, F$, where the optimal weights are
\begin{equation}
  w_{f, i}^{(j)} = \left( \sigma_f^{(j)} \right)^{-2} \left\{
    I(f) \sum_{j' = 1}^J \left( \sigma_f^{(j')} \right)^{-2}
  \right\}^{-1}.
  \label{eqn:mu_blue_w}
\end{equation}
\emph{One-sigma and two-sigma confidence intervals} for the unknown consensus mean $\mu_f$, are given by $\mu_f^* \pm S_f$ and $\mu_f^* \pm 2S_f$, respectively.

From \eqref{eqn:anova_underlying}, we also want to make inference on $Y_{f, i}$. Recall that $Y_{f, i} = \mu_f + \alpha_{f, i}$, where $\{ \alpha_{f, i} \}$ are random quantities that capture the departures from the longer-term climatology given by $\mu_f$, and they are assumed to have distribution $\Dist(0, \tau^2_f)$. Since the goal of the MIP is to infer a consensus for the underlying process $\{ Y_{f, i} \}$, BLUPs of these quantities are sought, along with their MSPEs. Straightforward use of the general Gauss--Markov Theorem \citep[e.g.,][]{rencherchristensen2012} results in the BLUP $Y_{f, i}^*$ and its MSPE, as follows:
\begin{align}
  Y_{f, i}^*
  & = \lambda_{f, i}^{(0)} \mu_f + \sum_{j = 1}^J \lambda_{f, i}^{(j)} Y_{f, i}^{(j)}
    \label{eqn:y_blup} \\
  \MSPE(Y_{f, i}^*)
  & \equiv E(Y_{f, i}^* - Y_{f, i})^2
    = \left( \tau_f^{-2} + \sum_{j' = 1}^J \left( \sigma_f^{(j')} \right)^{-2} \right)^{-1},
    \label{eqn:y_blup_mspe}
\end{align}
where (for the moment) all parameters, including $\mu_f$, are assumed known, and
\begin{equation}
  \lambda_{f, i}^{(j)} = \begin{cases}
    \tau_f^{-2} \left\{ \tau_f^{-2} + \sum_{j' = 1}^J \left( \sigma_f^{(j')} \right)^{-2} \right\}^{-1}; & j = 0, \\
    \left( \sigma_f^{(j)} \right)^{-2} \left\{ \tau_f^{-2} + \sum_{j' = 1}^J \left( \sigma_f^{(j')} \right)^{-2} \right\}^{-1}; & j = 1, \ldots, J. \\
  \end{cases}
  \label{eqn:y_blup_lambda}
\end{equation}
\emph{One-sigma and two-sigma prediction intervals} for $Y_{f, i}$ are given by $Y_{f, i}^* \pm \sqrt{\MSPE(Y_{f, i}^*)}$ and $Y_{f, i}^* \pm 2 \sqrt{\MSPE(Y_{f, i}^*)}$, respectively.

In the next section, we give likelihood-based estimates of the model parameters, which we notate as $\hat{\mu}_f$, $\{ (\hat{\sigma}_f^{(j)})^2 : j = 1, \ldots, J \}$, and $\hat{\tau}_f^2$. Then the \emph{empirical BLUE/BLUP} (written as EBLUE/EBLUP) equations are just those in \eqref{eqn:mu_blue}--\eqref{eqn:y_blup_lambda} with the model parameters, $\mu_f$, $\{ (\sigma_f^{(j)})^2 : j = 1, \ldots, J \}$, and $\tau_f^2$, \emph{replaced by their estimated quantities}. This yields the EBLUE for $\mu_f$, which we denote by $\hat{\mu}_f$, and an estimated variance denoted by $\hat{S}_f^2$. Further, it yields an EBLUP for $Y_{f, i}$, which we denote by $\hat{Y}_{f, i}$, and an estimated mean-squared prediction error, denoted by $\widehat{\MSPE}(\hat{Y}_{f, i})$. These formulas complete the specification of SUPE-ANOVA.

The likelihood-based estimates given in the next section require specification of statistical distributions and, to do this, we assume $\Dist(\cdot, \cdot)$ is $\Gau(\cdot, \cdot)$ for all distributions. When we assume Gaussianity, it should be noted that \eqref{eqn:y_blup} is best among all possible predictors, not just linear ones \citep[e.g.,][Sec.~4.1.2]{cressiewikle2011}.

\section{The initialization step: Estimation of parameters}
\label{sec:estimation}

Estimation of the parameters of the ANOVA model given by \eqref{eqn:anova_team} and \eqref{eqn:anova_underlying} is performed under the assumption that all unknown distributions are Gaussian, which yields a likelihood function of the model's parameters. The correlation parameters $\{ \rho_f^{(j, j')} \}$ can be modeled and estimated, but they are assumed to be zero in the application in Section~\ref{sec:oco2_mip}; here we omit them from the estimation equations, but see Appendix~\ref{sec:general_equations} where they are included. For the remaining parameters, it is known that the maximum likelihood estimates of variance parameters are biased and, to avoid this, we use restricted maximum likelihood (REML) estimation \citep{pattersonthompson1971,harville1977}. REML estimation maximizes the restricted likelihood, denoted by $L^{(r)}(\tauvec^2, \sigmavec^2 \mid \Yvec)$, where $\Yvec \equiv (Y_{1, 1}^{(1)}, Y_{1, 1}^{(2)}, \ldots, Y_{F, I(F)}^{(J)})'$ is the $(J \cdot \sum_{f = 1}^F I(f))$-dimensional vector of all outputs from the MIP, $\sigmavec^2 \equiv ((\sigma_1^{(1)})^2, (\sigma_1^{(2)})^2, \ldots, (\sigma_F^{(J)})^2)'$ is an $(F \cdot J)$-dimensional vector, and $\tauvec^2 = ( \tau_1^2, \ldots, \tau_F^2 )'$ is an $F$-dimensional vector. The definition of the restricted likelihood and a derivation of its analytical expression are given in Appendix~\ref{sec:general_equations}.

Small sample sizes can lead to variance estimates that need to be regularized in a manner similar to Twomey-Tikhonov regularization. This is achieved through the penalized restricted likelihood:
\begin{equation}
  L^{(p)}(\tauvec^2, \sigmavec^2 \mid \Yvec; \bvec)
  \equiv
  L^{(r)}(\tauvec^2, \sigmavec^2 \mid \Yvec) p(\sigmavec^2; \bvec),
  \label{eqn:penalized_restricted_likelihood}
\end{equation}
for
\[
  p(\sigmavec^2; \bvec) \equiv \prod_{f = 1}^F \prod_{j = 1}^J g_a\left( \left( \sigma_f^{(j)} \right)^2; b_f \right),
\]
where $\bvec \equiv (b_1, \ldots, b_F)'$ and $g_a(x; b) = x^{-a - 1}\exp(-b / x)$ for fixed $a > 0$. The function $g_a(x; b)$ is proportional to the density of an inverse gamma distribution with shape $a$ and scale $b$, which is a common choice when making Bayesian inference on variance parameters. The value of the scales $\bvec$ in the penalty $p(\sigmavec^2; \bvec)$ is considered unknown and is estimated, whereas the shape $a$ is fixed and pre-specified to guard against one team's output being over-weighted in the BLUEs/BLUPs. That is, for a given factor $f \in \{ 1, \ldots, F \}$, the penalty puts a soft limit on the ratio between the largest and the smallest team-specific variances, $\{ (\sigma_f^{(1)})^2, \ldots, (\sigma_f^{(J)})^2 \}$, to a degree specified by $a$, while the actual scale of the variances as a whole is left unknown and estimated through $b_f$.

Calibration of the shape $a$ can be achieved by interpreting the penalty as a Bayesian prior on the variances. In that case, we specify that the ratio of the prior 97.5th percentile of the variances to the prior 2.5th percentile of the variances is equal to four, which yields the fixed value of $a = 8.48$. This choice means that it is unlikely that the largest variance is more than four times larger than the smallest variance.

The expression $L^{(p)}(\tauvec^2, \sigmavec^2 \mid \Yvec; \bvec)$ in \eqref{eqn:penalized_restricted_likelihood} is maximized with respect to the model parameters $\tauvec^2$ and $\sigmavec^2$ and with respect to the penalization scale $\bvec$, to yield the penalized REML estimates $\{ (\hat{\sigma}_f^{(j)})^2 : f = 1, \ldots, F,\enskip j = 1, \ldots, J \}$ and $\{ \hat{\tau}_f^2 : f = 1, \ldots, F \}$. Appendix~\ref{sec:general_equations} gives the formulas for the general case where correlations are present. For $f = 1, \ldots, F$, these estimates are substituted into \eqref{eqn:mu_blue}, the BLUE for $\mu_f$, to give the EBLUE of $\mu_f$, denoted by $\hat{\mu}_f$, with empirical variance $\hat{S}^2_f$. For $f = 1, \ldots, F$, $i = 1, \ldots, I(f)$, the EBLUP of $Y_{f, i}$, denoted by $\hat{Y}_{f, i}$, with empirical MSPE, denoted by $\widehat{\MSPE}(\hat{Y}_{f, i})$, is obtained similarly.

The statistical methodology given in Sections~\ref{sec:statistical_anova}, \ref{sec:optimal_weighting}, and \ref{sec:estimation} is applied in the next section to a MIP comprising nine flux inversion--based estimates of global carbon dioxide fluxes.

\section{\texorpdfstring{The OCO-2 MIP: Consensus CO$_2$ fluxes}{The OCO-2 MIP: Consensus CO2 fluxes}}
\label{sec:oco2_mip}

The rise in the atmospheric concentration of the greenhouse gas carbon dioxide (CO$_2$) is the main determinant of global warming \citep{ipcc2013}, and estimation of the sources and sinks (called fluxes) of CO$_2$ is a high priority for the global community \citep{friedlingsteinetal2014}. Fluxes of CO$_2$ arise from both anthropogenic and natural processes, and the latter are hard to observe directly, as they occur over large spatial and temporal scales. One way to estimate the natural fluxes, called flux inversion, is through observation of their influence on atmospheric concentrations of CO$_2$. As natural fluxes are so difficult to measure otherwise, flux inversions for CO$_2$ play an important role in quantifying Earth's carbon budget.

Flux inversion is an ill-posed inverse problem, in which an atmospheric transport model is used to simulate the transport of the trace gas from its sources to the times and places of its observation. The transport model is combined with a procedure, usually based on a Bayesian state-space model, to work backwards from observations to fluxes. The data used by inversions are typically a combination of surface, aircraft, and ship observations, and data collected by spaceborne instruments. In this section, we shall use the output from a MIP based on CO$_2$ data from NASA's Orbiting Carbon Observatory-2 (OCO-2) satellite \citep{crowelletal2019} to obtain consensus flux estimates and predictions.

Flux inversion is subject to many forms of model uncertainty, which complicate interpretation of the estimated fluxes. These uncertainties stem from questions about the fidelity of simulated atmospheric transport, about how best to parameterize the unknown fluxes, and about the choice of prior for the flux field, among others. To address these in a rigorous manner, the OCO-2 project established a series of MIPs where an ensemble of flux outputs was assembled from submissions by different teams, who performed global flux inversions under a common protocol. The MIPs involve the use of OCO-2 retrievals of the vertical column--average CO$_2$ concentration, as well as flask and \emph{in situ} observations of CO$_2$ concentrations.

The first of the MIPs concluded in 2019, and its results were reported by \citet{crowelletal2019}. The satellite data in this MIP were based on version 7r of the OCO-2 data, so henceforth we refer to it as MIPv7. Nine teams participated in MIPv7. The results were released as a set of aggregated monthly non-fossil-fuel fluxes spanning 27 disjoint, large regions derived from TransCom3 regions, shown in Figure~\ref{fig:region_map} (where the unlabeled areas are assumed to have zero flux) and listed in Table~\ref{tab:region_table}. The outputs from the MIP covered each month over the two years from January 2015 to December 2016, inclusive. Each team was tasked with producing estimated fluxes from three observation types: from an inversion based on flask and \emph{in situ} measurements only (IS), from an inversion based on OCO-2 retrievals made in land glint (LG) mode, and from an inversion based on OCO-2 retrievals made in land nadir (LN) mode.

To apply the SUPE-ANOVA framework to MIPv7, the fluxes from each mode were considered separately. Then, the factor of interest is $f = (s, r)$, where $s = 1, \ldots, 4$ is a season (starting with December-January-February [DJF]), and $r = 1, \ldots, 27$ is a spatial region. The replicates $i = 1, \ldots, 6$ are the six months within a season for the two-year study period, so $I(f) = 6$ for all factors. From \eqref{eqn:anova_team} and \eqref{eqn:anova_underlying}, the flux output of the $j$th team is modeled by,
\begin{equation}
  Y_{s, r, i}^{(j)}
  = Y_{s, r, i} + \eta_{s, r, i}^{(j)}, \quad \eta_{s, r, i}^{(j)} \sim \Gau\left( 0, \left( \sigma_{s, r}^{(j)} \right)^2 \right),
  \label{eqn:mip_model1}
\end{equation}
for $j = 1, \ldots, 9$, where $\{ \eta_{s, r, i}^{(j)} : j = 1, \ldots, 9 \}$ are uncorrelated; we assume they are uncorrelated for simplicity, but, if desired, it is relatively easy to incorporate correlations using the equations in Appendix~\ref{sec:general_equations}. The consensus flux is modeled by
\begin{equation}
  Y_{s, r, i}
  = \mu_{s, r} + \alpha_{s, r, i}, \quad \alpha_{s, r, i} \sim \Gau(0, \tau_{s, r}^2).
  \label{eqn:mip_model2}
\end{equation}
The term $Y_{s, r, i}$ represents the MIP consensus for the underlying geophysical flux for month $i$ in season $s$ and region $r$; $\mu_{s, r}$ is the more-slowly-varying consensus climatological flux for the season and region; $\tau_{s, r}^2$ is the variance of the deviations of the monthly fluxes from the climatological flux; and $(\sigma_{s, r}^{(j)})^2$ is the variance of the $j$th team's flux output within each region and season.

The six replicates within each factor $(s, r)$ are too few for reliable estimation of the variance parameters, but this can be addressed straightforwardly by grouping regions with \emph{similar variability} within a season. Hence, for the four seasons $s = 1, \ldots, 4$, variance parameters $\tau_{s, r}^2$, $\{ (\sigma_{s, r}^{(j)})^2 : j = 1, \ldots, J \}$, and the penalization scales $b_{s, r}$ were constrained to have identical values within grouped regions. We present a clustering approach to construct groups in Appendix~\ref{sec:clustering_procedure}; the chosen groups are shown in Figure~\ref{fig:clustering_map}.

The parameters of the model in \eqref{eqn:mip_model1} and \eqref{eqn:mip_model2} were estimated according to the initialization step given in Section~\ref{sec:estimation}, and the SUPE-ANOVA framework was used to obtain the consensus flux predictions $\{ \hat{Y}_{s, r, i} : s = 1, \ldots, 4;\enskip r = 1, \ldots, 27;\enskip i = 1, \ldots, 6 \}$. To check whether the model has good fit to the fluxes, we define the standardized errors, $\hat{\eta}_{s, r, i}^{(j)} = (Y_{s, r, i}^{(j)} - \hat{Y}_{s, r, i}) / \hat{\sigma}_{s, r}^{(j)}$, for $s = 1, \ldots, 4$, $r = 1, \ldots, 27$, $i = 1, \ldots, 6$, and $j = 1, \ldots, 9$. Under the Gaussian assumptions used to fit the model, the standardized errors $\{ \hat{\eta}_{s, r, i}^{(j)} \}$ should be approximately $\text{i.i.d.}\,\Gau(0, 1)$. We diagnosed this visually using Q-Q plots, by plotting $\{ \hat{\eta}_{s, r, i}^{(j)} \}$ for each season and group of regions against the theoretical quantiles of the $\Gau(0, 1)$ distribution; under the assumptions, the relationship should be approximately on a 45$\degree$ line. Figure~\ref{fig:qqplots} shows these Q-Q plots, which do not indicate any obvious violations of the model assumptions.

\begin{figure}
  \begin{center}
    \includegraphics{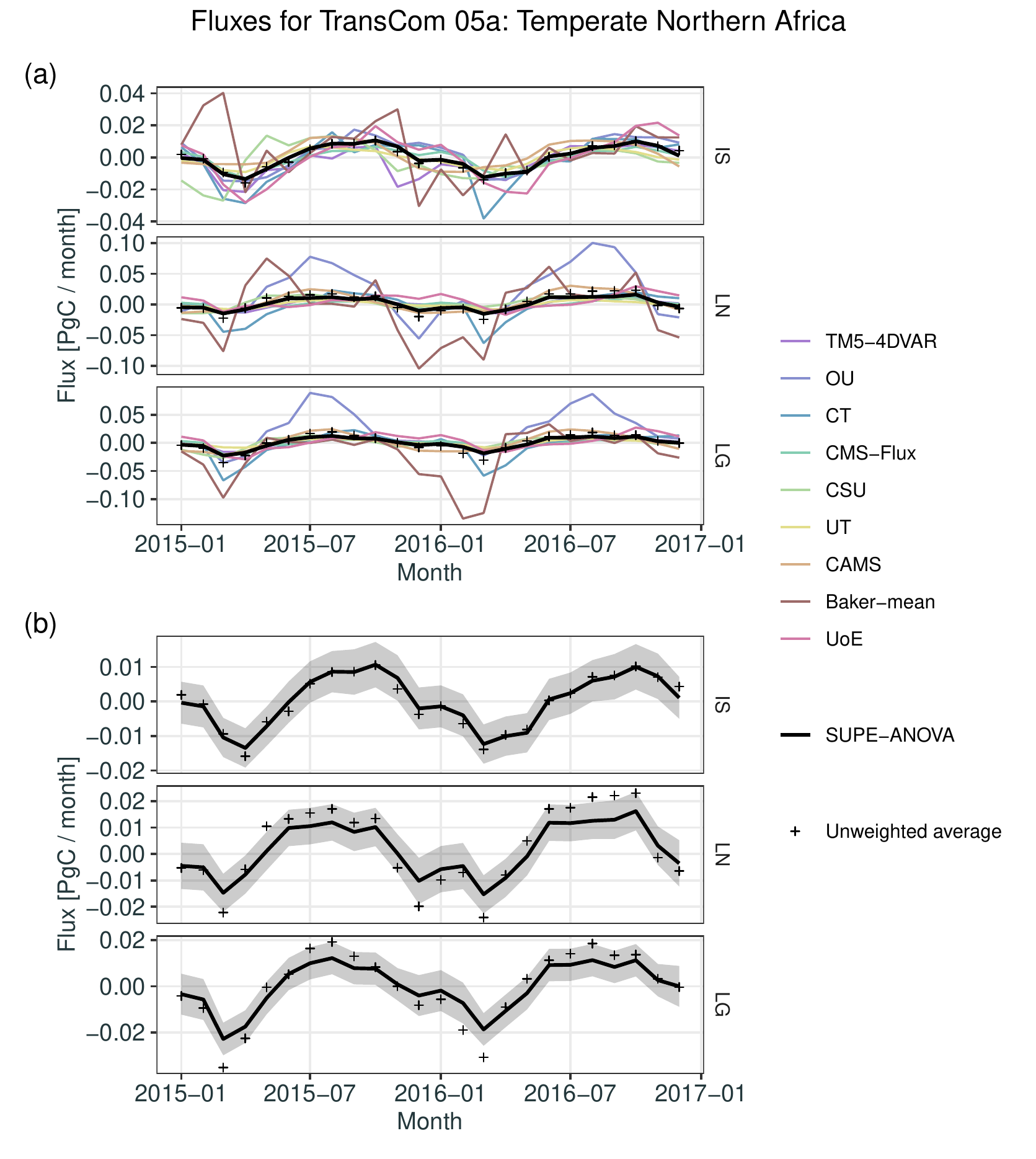}
  \end{center}

  \caption{
    SUPE-ANOVA consensus (black lines) and unweighted-average (black crosses) non-fossil-fuel fluxes for the region T05a (Temperate Northern Africa). (a) The nine individual OCO-2 MIPv7 fluxes are shown with colored lines. (b) The individual fluxes are omitted, and a shaded area is added to show the 95\% prediction interval around the SUPE-ANOVA consensus fluxes.
  }
  \label{fig:flux_regional}
\end{figure}

As an example of predicting consensus fluxes at the regional level, Figure~\ref{fig:flux_regional}(a) shows the SUPE-ANOVA predictions as a time series for one region, T05a (Temperate Northern Africa), as well as the time series of fluxes from each of the nine MIP participants; the unweighted averages, $\bar{Y}_{s, r, i} = \frac{1}{9} \sum_{j = 1}^9 Y_{s, r, i}^{(j)}$, are also shown. Separate panels give the time series for each observation type (IS, LN, and LG). Figure~\ref{fig:flux_regional}(b) repeats Figure~\ref{fig:flux_regional}(a) but with the participating teams' fluxes omitted to allow the vertical axis to be re-scaled. It also shows 95\% prediction intervals for the predicted consensus fluxes, where the upper and lower values are $\hat{Y}_{s, r, i} \pm 1.96 \times \sqrt{\widehat{\MSPE}(\hat{Y}_{s, r, i})}$. For the IS fluxes in the region T05a, the predicted consensus hardly differs from the unweighted average. By contrast, for the LN and LG fluxes, the predicted consensus sometimes differs significantly from the unweighted average.

\begin{figure}
  \begin{center}
    \includegraphics{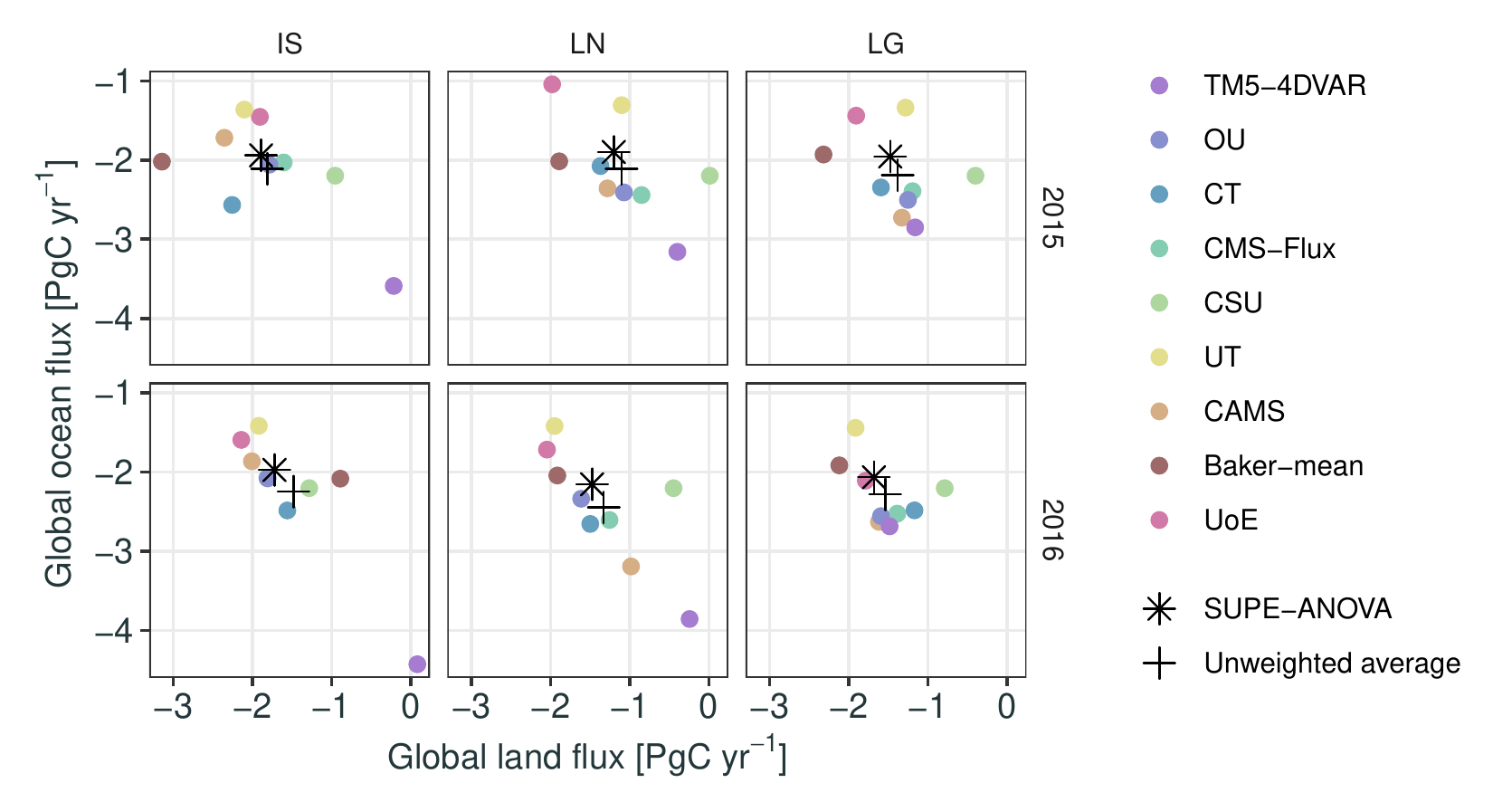}
  \end{center}

  \caption{
    SUPE-ANOVA consensus ($\ast$) and unweighted-average (+) non-fossil-fuel OCO-2 MIPv7 annual global ocean fluxes versus annual global land fluxes for each observation type (IS, LN, and LG) and year (2015 and 2016). Colored points show the underlying MIP outputs from the nine teams.
  }
  \label{fig:flux_aggregates_bivariate}
\end{figure}

The SUPE-ANOVA predictions of the consensus fluxes and the unweighted-average fluxes at the regional level were also aggregated spatially and temporally to yield the annual global land and annual global ocean fluxes. These are shown in Figure~\ref{fig:flux_aggregates_bivariate}, which also shows the land--ocean aggregates for the nine teams in the MIP. Fluxes are given for all three observation types, IS, LN, and LG, and for both study years, 2015 and 2016. The SUPE-ANOVA consensus fluxes imply a larger land sink and a smaller ocean sink than do the unweighted averages. Also, for the IS observation type, there are more extreme values from the \nospellcheck{TM5-4DVAR} team then the other teams, and these have been downweighted in the consensus estimate, relative to the weights they received in the unweighted average.

\begin{figure}
  \begin{center}
    \includegraphics{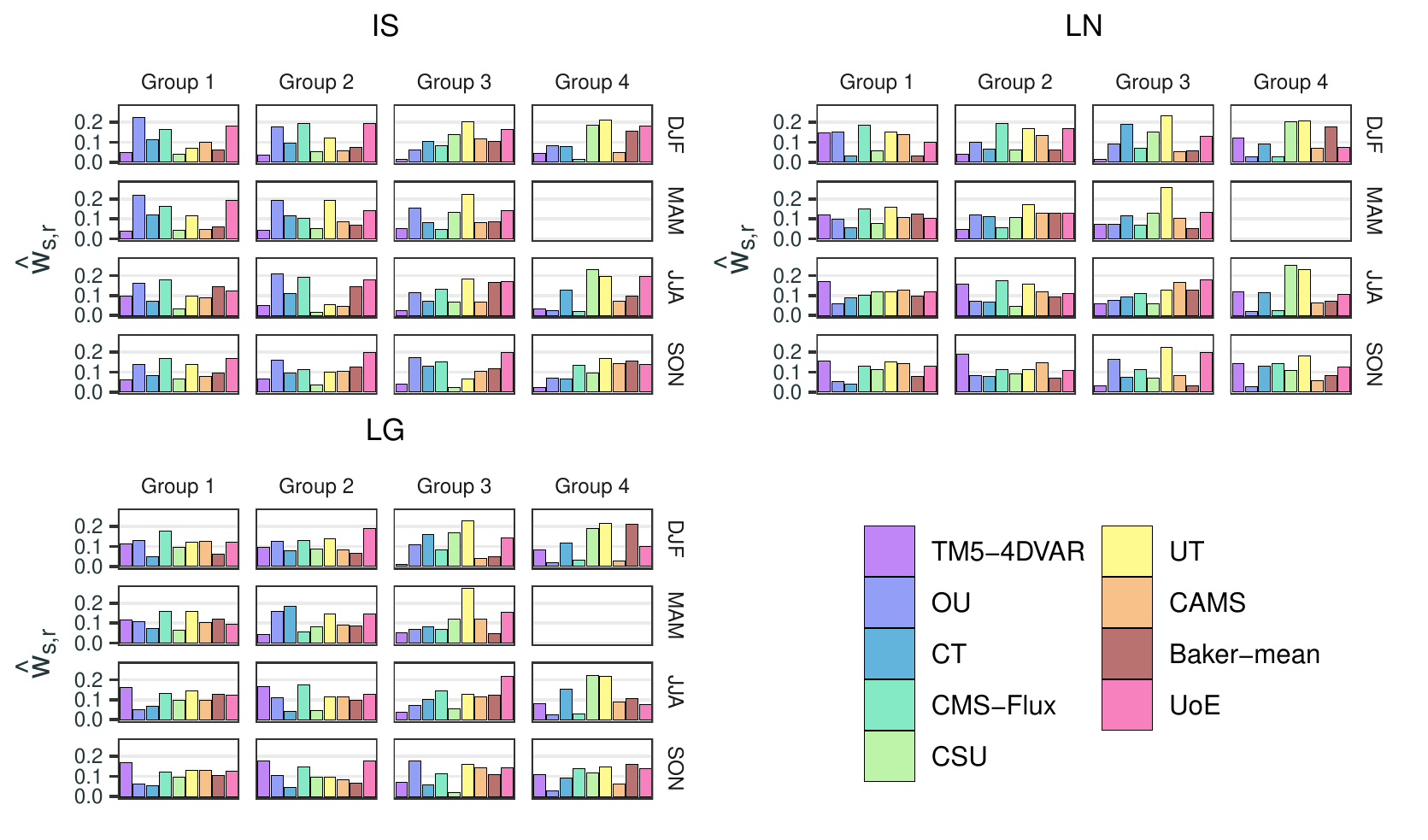}
  \end{center}

  \caption{
    Estimated weights, $\{ \hat{w}_{s, r}^{(j)} : s = 1, \ldots, 4; r = 1, \ldots, 27; j = 1, \ldots, 9 \}$, obtained from SUPE-ANOVA, for each observation type (IS, LN, and LG), season ($s)$, region ($r$, shown according to its group; see Figure~\ref{fig:clustering_map}), and team ($j$) in the OCO-2 MIPv7.
  }
  \label{fig:climatological_weights}
\end{figure}

The different teams' weights used in the SUPE-ANOVA framework vary between seasons and regions, although they are identical within the cluster groups. For a given $(s, r)$, define the climatological weights, $\hat{w}_{s, r}^{(j)} = (\hat{\sigma}_{s, r}^{(j)})^{-2} / \left[ \sum_{j' = 1}^J (\hat{\sigma}_{s, r}^{(j')})^{-2} \right]$, for $j = 1, \ldots, 9$. From \eqref{eqn:mu_blue}, these weights correspond to the relative weighting of each participant in estimating the consensus mean climatological flux, $\mu_{s, r}$. Figure~\ref{fig:climatological_weights} shows the climatological weights for the three observation types (IS, LN, LG), for each season and grouping of regions within each season. From inspection of the weights in Figure~\ref{fig:climatological_weights}, the UT team and the \nospellcheck{UoE} team generally have higher weights due to their associated estimated variances being consistently smaller.

\section{Discussion and conclusions}
\label{sec:discussion}

A reader of this article might ask whether in the North American temperature example given in Section~\ref{sec:optimal_weighting}, the simple unweighted average of the MIP outputs provides a reasonable estimate of $\mu_\Tmp$. Under the SUPE-ANOVA framework, that estimate $\bar{\mu}_\Tmp$ given by Equation~\eqref{eqn:nat_homo_blue} is unbiased. However, it is not a statistically efficient estimator under our consensus model. This can have a profound effect on the design of the MIP. Using the optimal weights, namely, $w_\Tmp^{(j)}$ inversely proportional to $(\sigma_\Tmp^{(j)})^2 \equiv \var(Y_\Tmp^{(j)})$ for $j = 1, \ldots, J$, means that the precision of $\mu_\Tmp^* = \sum_{j = 1}^J w_\Tmp^{(j)} Y_\Tmp^{(j)}$ can be matched to the precision of $\bar{\mu}_\Tmp$ but with $J$ (much) less than 20. Since $\mu_\Tmp^*$ is also unbiased, this optimally weighted mean has a distinct inferential advantage over the unweighted mean, $\bar{\mu}_\Tmp$, when both are based on the same $J = 20$ outputs $\{ Y_\Tmp^{(1)}, \ldots, Y_\Tmp^{(20)} \}$. Furthermore, the simple variance, $\sigma^2_\Tmp / 20$, in Equation~\eqref{eqn:nat_homo_blue} is an incorrect expression for $\var(\bar{\mu}_\Tmp)$ when it is no longer true that $(\sigma_\Tmp^{(1)})^2 = \ldots = (\sigma_\Tmp^{(20)})^2 = \sigma_\Tmp^2$. The correct expression is
\[
  \var(\bar{\mu}_\Tmp) = \sum_{j = 1}^{20} \left( w_\Tmp^{(j)} \right)^2 \left( \sigma_\Tmp^{(j)} \right)^2,
\]
which will give valid consensus inferences and correct coverage for the one-sigma and two-sigma intervals presented in Section~\ref{sec:optimal_weighting}, albeit wider than the intervals based on $\var(\mu_\Tmp^*)$.

A number of articles reviewed in Section~\ref{sec:introduction} have taken a Bayesian viewpoint \citep{tebaldietal2005,manningetal2009,smithetal2009,tebaldisanso2009,chandler2013}. A Bayesian spatial ANOVA was used in \citet{sainetal2011}, \citet{kangetal2012}, and \citet{kangcressie2013} to analyze projected temperature in a MIP called the North American Regional Climate Change Assessment Program \citep[NARCCAP; see][]{mearnsetal2012}. This approach adds another layer of complexity to consensus inference in a MIP, where prior distributions on mean and variance parameters need to be specified, sensitivity to the specification assessed, and posterior-distribution sampling algorithms developed. In contrast, our SUPE-ANOVA framework does not depend on priors, is fast to compute, and results in easy-to-interpret weights on the $J$ MIP outputs within the ANOVA factors (e.g., season, region).

We return to the general SUPE-ANOVA framework developed in Sections~\ref{sec:statistical_anova} and \ref{sec:optimal_weighting} to show how the optimal weights could be used in conjunction with verification data. Let $\{ Z_k : k = 1, \ldots, K \}$ denote the verification data, with associated measurement-error variances $\{ \var(Z_k) : k = 1, \ldots, K \}$. Each datum $Z_k$ can be indexed according to which factor $f$ and replicate $i \in \{ 1, \ldots, I(f) \}$ it belongs. Hence, rewrite $Z_k$ as $Z_{f(k), i(k)}$, which will be compared to the $J$ outputs, $\{ Y_{f(k), i(k)}^{(j)} : j = 1, \ldots, J \}$. Then a simple and statistically justifiable way to assign weights to each team's output is to define
\begin{equation}
  \nu^{(j)}
  \propto
    \left(
      \prod_{f = 1}^F \prod_{i = 1}^{I(f)}
      w_{f, i}^{(j)}
    \right)
    \left\{
      \prod_{k = 1}^K
      \exp \left[
        -\frac{
          \left( Z_{f(k), i(k)} - Y_{f(k), i(k)}^{(j)} \right)^2
        }{
          2\,\var(Z_{f(k), i(k)})
        }
      \right]
    \right\},
  \label{eqn:bayesian_model_averaging}
\end{equation}
where $\sum_{j = 1}^J \nu^{(j)} = 1$. Equation~\eqref{eqn:bayesian_model_averaging} shows how the SUPE-ANOVA weights, $\{ w_{f, i}^{(j)} \}$, can be used to define a prior in a Bayesian Model Averaging (BMA) scheme, where $\{ \nu^{(j)} : j = 1, \ldots, J \}$ can be interpreted as the posterior weights assigned to the $J$ models, given the verification data $\{ Z_k : k = 1, \ldots K \}$. That is, the $J$ models are weighted based on verification data without making the common `truth-plus-error' assumption \citep{bishopabramowitz2013}, or even without making the more general assumption of co-exchangeability between the MIP outputs and the data \citep{rougieretal2013}. Now, since the BMA weights have lost their dependence on the $F$ factors, they will not yield optimal estimates of $\mu_f$, or optimal predictions of $\{ Y_{f, i} \}$; instead, they can be used to suggest which model is the most reliable overall in the sense of REA.

In summary, basic statistical principles have been followed to make optimal consensus inference in a MIP. Once the variance/covariance parameters in the SUPE-ANOVA framework have been estimated in an initialization step, the weights are immediate; code in the R programming language \citep{R_Core_2020} is available at \url{https://github.com/mbertolacci/supe-anova-paper} for the OCO-2 MIP analyzed in Section~\ref{sec:oco2_mip}.


\section*{Acknowledgements}

This project was largely supported by the Australian Research Council (ARC) Discovery Project (DP) DP190100180 and NASA ROSES grant \nospellcheck{NNH20ZDA001N-OCOST}. Noel \nospellcheck{Cressie's} research was also supported by ARC DP150104576, and Andrew \nospellcheck{Zammit-Mangion's} research was also supported by an ARC Discovery Early Career Research Award (\nospellcheck{DECRA}) DE180100203. The authors are grateful to the OCO-2 Flux Group for their feedback and suggestions.

\section*{Data Availability Statement}

The data for the OCO-2 MIP analyzed in Section~\ref{sec:oco2_mip} are described by \citet{crowelletal2019} and may be found on the OCO-2 MIPv7 website hosted by the \nospellcheck{NOAA} Global Monitoring Laboratory (\url{https://gml.noaa.gov/ccgg/OCO2/}).

\bibliography{references}

\section*{Appendices}
\appendix

\renewcommand\thefigure{\thesection\arabic{figure}}
\renewcommand\thetable{\thesection\arabic{table}}
\setcounter{figure}{0}
\setcounter{table}{0}

\section{Additional Figures and Tables}
\label{sec:appendix_figures}

\begin{figure}[H]
  \begin{center}
    \includegraphics{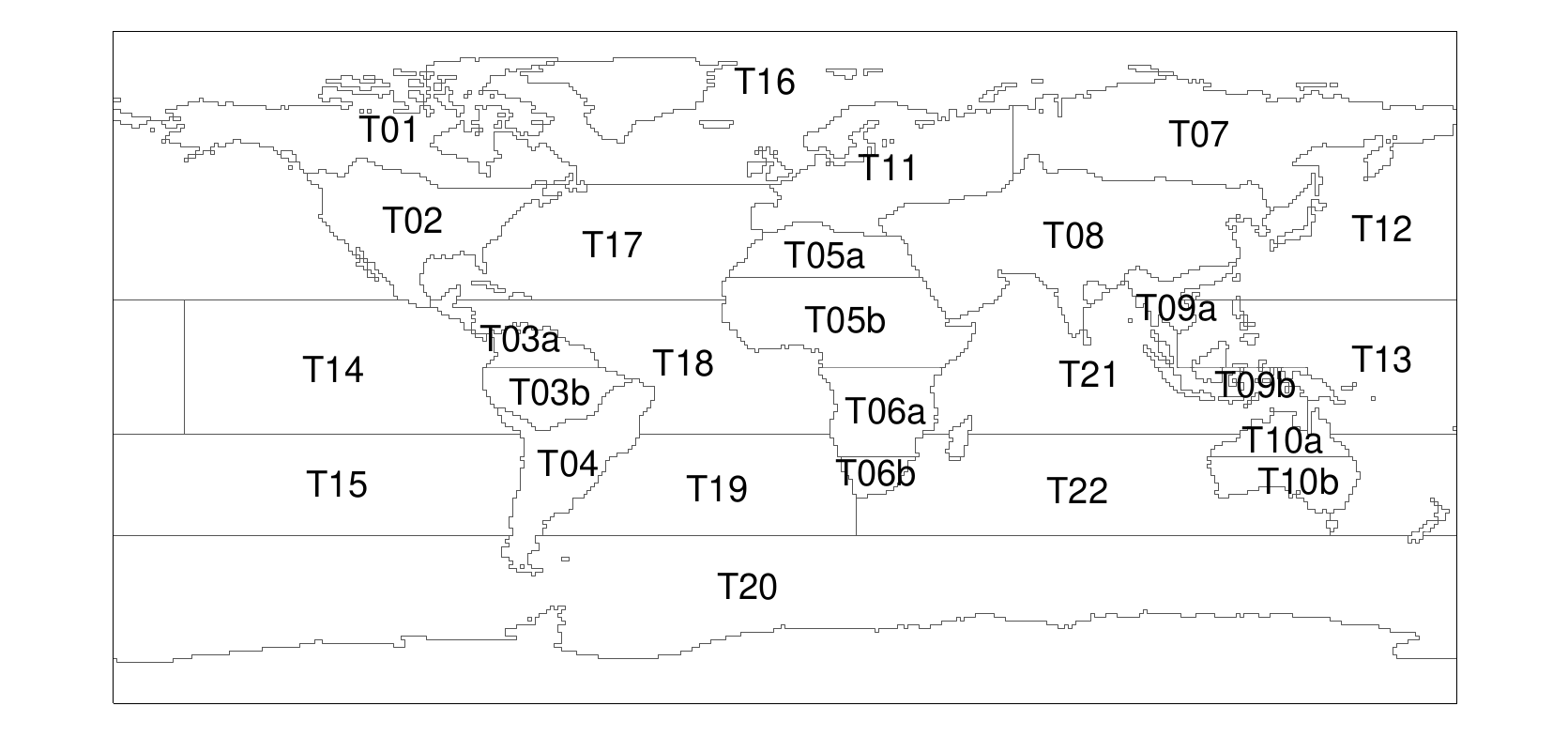}
  \end{center}

  \caption{
    The spatial extents of the 27 regions (derived from the TransCom3 regions) over which the OCO-2 MIPv7 fluxes are estimated. The names of the regions are given in Table~\ref{tab:region_table}.
  }
  \label{fig:region_map}
\end{figure}

\begin{table}[H]
  \begin{center}
    \singlespacing
    \begin{tabular}{l|l||l|l}
\textbf{Code} & \textbf{Name} & \textbf{Code} & \textbf{Name} \\ \hline
T01 & North American Boreal & T10b & Temperate Australia \\
T02 & North American Temperate & T11 & Europe \\
T03a & Northern Tropical South America & T12 & North Pacific Temperate \\
T03b & Southern Tropical South America & T13 & West Pacific Tropical \\
T04 & South American Temperate & T14 & East Pacific Tropical \\
T05a & Temperate Northern Africa & T15 & South Pacific Temperate \\
T05b & Northern Tropical Africa & T16 & Northern Ocean \\
T06a & Southern Tropical Africa & T17 & North Atlantic Temperate \\
T06b & Temperate Southern Africa & T18 & Atlantic Tropical \\
T07 & Eurasia Boreal & T19 & South Atlantic Temperate \\
T08 & Eurasia Temperate & T20 & Southern Ocean \\
T09a & Northern Tropical Asia & T21 & Indian Tropical \\
T09b & Southern Tropical Asia & T22 & South Indian Temperate \\
T10a & Tropical Australia &  &  \\
\end{tabular}

  \end{center}
  \caption{
    The 27 regions over which the OCO-2 MIPv7 fluxes are estimated \citep{crowelletal2019}. Each region has a code and a name. The locations of the regions are shown in Figure~\ref{fig:region_map}.
  }
  \label{tab:region_table}
\end{table}

\begin{figure}[H]
  \begin{center}
    \includegraphics{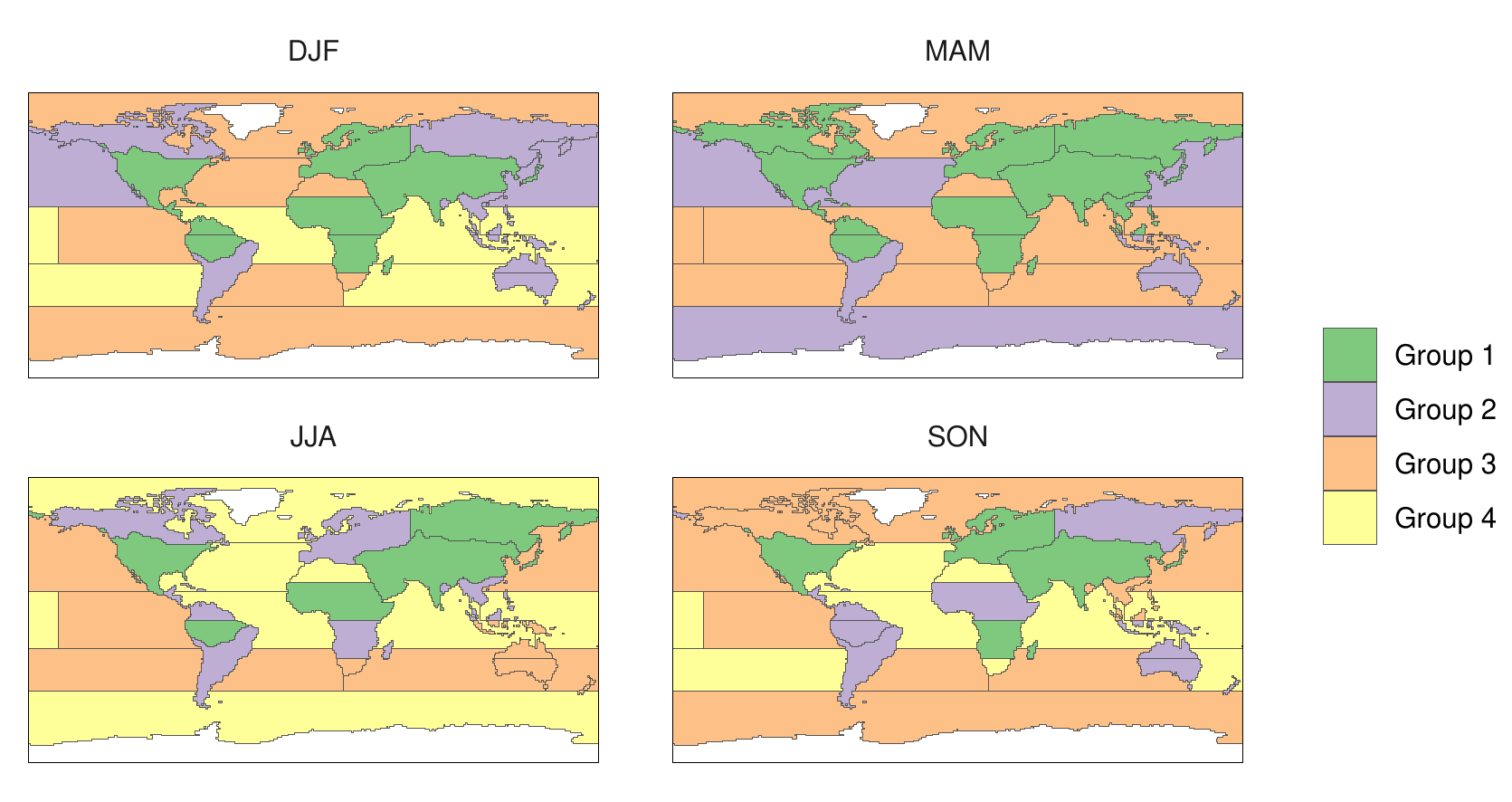}
  \end{center}

  \caption{
    The groups chosen using the clustering procedure for the OCO-2 MIPv7 for each of the four seasons, DJF, MAM, JJA, and SON. The group numbers are ordered within each season, from the clustered regions exhibiting the most variability, group one, to those exhibiting the least variability, group four (or group three for MAM, since it contains only three clusters).
  }
  \label{fig:clustering_map}
\end{figure}

\begin{figure}[H]
  \begin{center}
    \includegraphics{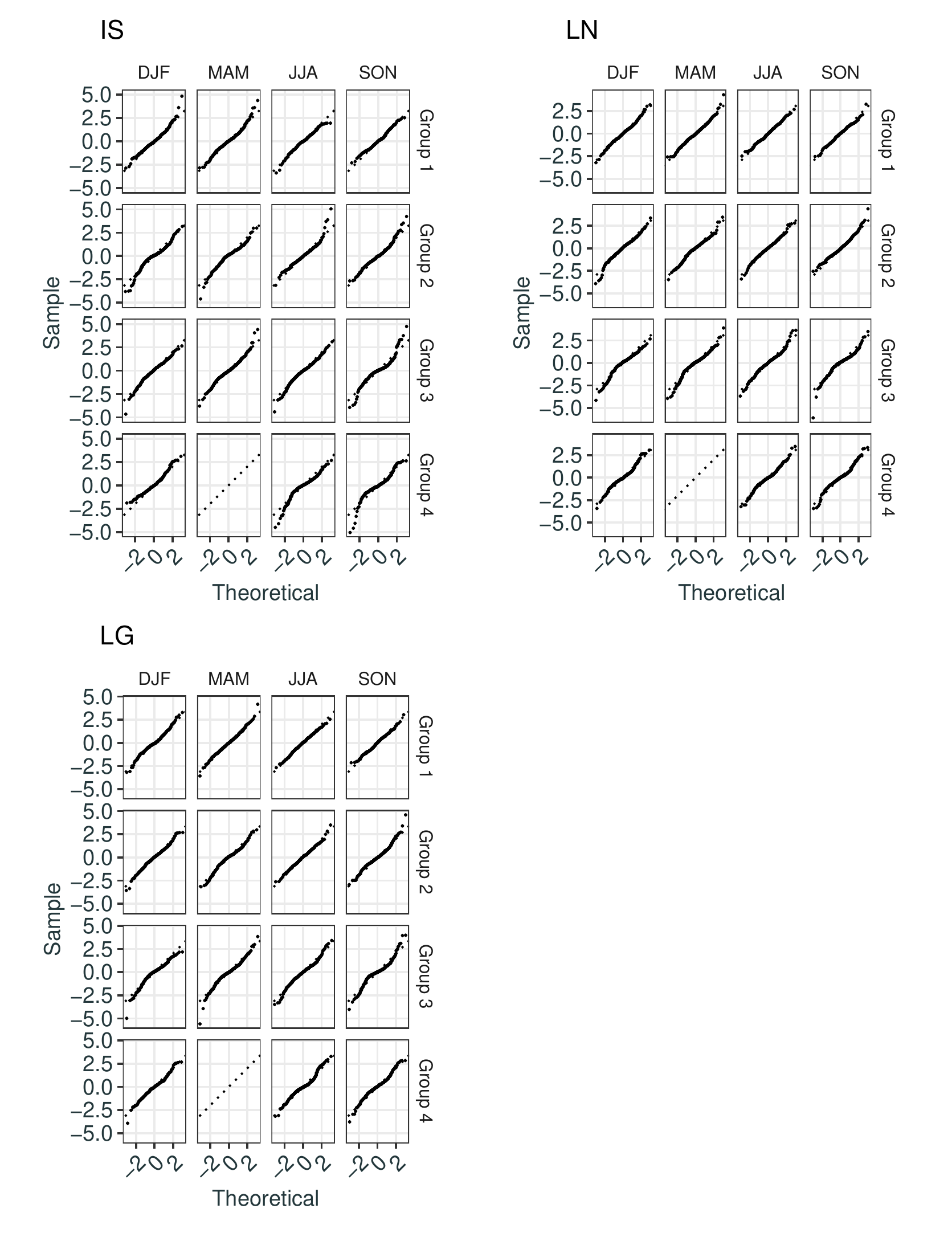}
  \end{center}

  \caption{
    Q-Q plots showing the standardized residuals, $\{ \hat{\eta}_{s, r, i}^{(j)} : s = 1, \ldots, 4, r = 1, \ldots, 27, i = 1, \ldots, 6, j = 1, \ldots, 9 \}$, versus the theoretical quantiles of the $\Gau(0, 1)$ distribution for the SUPE-ANOVA model applied to the OCO-2 MIPv7.
  }
  \label{fig:qqplots}
\end{figure}

\section{The general SUPE-ANOVA equations and the restricted likelihood}
\label{sec:general_equations}

To derive the general estimation/prediction equations and the restricted likelihood of the ANOVA model given by \eqref{eqn:anova_team} and \eqref{eqn:anova_underlying}, we first rewrite the model in matrix-vector form as
\begin{equation}
  \Yvec = \Xmat \muvec + \Zmat \alphavec + \etavec,
  \label{eqn:model_vec}
\end{equation}
where $\etavec \equiv (\eta_{1, 1}^{(1)}, \eta_{1, 1}^{(2)}, \ldots, \eta_{F, I(F)}^{(J)})'$ is a $(J \cdot \sum_{f = 1}^F I(f))$-dimensional vector; $\muvec \equiv (\mu_1, \ldots, \mu_F)'$ is an $F$-dimensional vector; $\alphavec \equiv (\alpha_{1, 1}, \alpha_{1, 2}, \ldots, \alpha_{F, I(F)})'$ is a $(\sum_{f = 1}^F I(f))$-dimensional vector; and $\Xmat$ and $\Zmat$ are matrices of ones and zeros that map the entries of $\muvec$ and $\alphavec$, respectively, to the appropriate entries of $\Yvec$. Let $\Sigmamat_\eta \equiv \var(\etavec)$ and $\Sigmamat_\alpha \equiv \var(\alphavec)$. The matrix $\Sigmamat_\alpha$ is diagonal with the entries of $\tauvec^2$ on its diagonal, while $\Sigmamat_\eta$ has the entries of $\sigmavec^2$ on its diagonal and $\{ \rho_f^{(j, j')} \sigma_f^{(j)} \sigma_f^{(j')} : j \neq j' \}$ in its off-diagonal entries. For both matrices, the entries are repeated as needed for each replicate $i = 1, \ldots, I(f)$. Under \eqref{eqn:model_vec}, the marginal mean and variance of $\Yvec$ are, therefore, $E(\Yvec) = \Xmat \muvec$ and $\Sigmamat_Y \equiv \var(\Yvec) = \Zmat \Sigmamat_\alpha \Zmat' + \Sigmamat_\eta$, respectively.

The multivariate Gauss--Markov theorem \citep[e.g.,][]{rencherchristensen2012} yields the BLUE under \eqref{eqn:model_vec} as
\begin{equation}
  \muvec^* = (\Xmat' \Sigmamat_Y^{-1} \Xmat)^{-1} \Xmat' \Sigmamat_Y^{-1} \Yvec.
  \label{eqn:mu_blue_general}
\end{equation}
The variance of $\muvec^*$ is given by $\Sigmamat_\mu \equiv (\Xmat' \Sigmamat_Y^{-1} \Xmat)^{-1}$. The BLUP for $\alphavec$ is then
\begin{equation}
  \alphavec^* = \Sigmamat_\alpha \Zmat' \Sigmamat_Y^{-1} (\Yvec - \Xmat \muvec^*),
\end{equation}
with prediction variance $\Sigmamat_\alpha^* \equiv \Sigmamat_\alpha - \Sigmamat_\alpha \Zmat' \Sigmamat_Y^{-1} \Zmat \Sigmamat_\alpha$. The BLUP and prediction variance for $\Yvec$ follow as $\Xmat \muvec^* + \Zmat \alphavec^*$ and $\Zmat \Sigmamat_\alpha^* \Zmat'$, respectively. When $\rho_f^{(j, j')} = 0$ for $f = 1, \ldots, F$ and $j, j' = 1, \ldots, J$, the vector expressions for the BLUE and the BLUP given here can be shown to be equivalent to the element-wise expressions \eqref{eqn:mu_blue} and \eqref{eqn:y_blup}, respectively.

To derive the restricted likelihood, we first assume that all unknown distributions are Gaussian. Under this assumption, the distribution of $\Yvec$ is multivariate normal with $\Yvec \sim \Gau(\Xmat \muvec, \Sigmamat_Y)$. Viewed as a function of its parameters, $\muvec, \tauvec^2$, $\sigmavec^2$, and $\rhovec \equiv (\rho_1^{(1, 2)}, \rho_1^{(1, 3)}, \ldots, \rho_F^{(J - 1, J)})'$, where $\tauvec^2$ and $\sigmavec^2$ are defined in Section~\ref{sec:estimation}, the density of this distribution is the likelihood $L(\muvec, \tauvec^2, \sigmavec^2, \rhovec \mid \Yvec)$. The restricted likelihood is equal to the integral of the likelihood with respect to $\muvec$ \citep{harville1977}; that is,
\begin{align*}
  L^{(r)}(
    \tauvec^2,
    \sigmavec^2,
    \rhovec
    \mid
    \Yvec
  )
  & \equiv
    \int L(
      \muvec,
      \tauvec^2,
      \sigmavec^2,
      \rhovec
      \mid
      \Yvec
    )\ \mathrm{d}\muvec\\
  & \propto
    |\Sigmamat_\mu|^{1 / 2} |\Sigmamat_Y|^{-1 / 2} \exp\left\{
    -\frac{1}{2}(\Yvec - \muvec^*)' \Sigmamat_Y^{-1} (\Yvec - \muvec^*)
    \right\},
\end{align*}
where the constant of proportionality does not depend on the unknown parameters. Where $\rhovec$ is assumed to be known, as in Sections~\ref{sec:estimation} and \ref{sec:oco2_mip}, we write the restricted likelihood as $L^{(r)}(\tauvec^2, \sigmavec^2 \mid \Yvec)$.

\section{Clustering procedure to group regions of similar variability}
\label{sec:clustering_procedure}

For the OCO-2 MIPv7 example in Section~\ref{sec:oco2_mip}, we established the groupings of regions with similar variability as follows: Let $\tilde{S}_{s, r}$ be the empirical median of the set $\{ |Y_{s, r, i}^{(j)} - \tilde{Y}_{s, r, i}| : i = 1, \ldots, 6;\enskip j = 1, \ldots, 9 \}$, where $\tilde{Y}_{s, r, i}$ is the empirical median of the set $\{ Y_{s, r, i}^{(j)} : j = 1, \ldots, 9 \}$. The value $\tilde{S}_{s, r}$ is a robust empirical estimate of the ensemble variability in the season and region. We then divided the 27 regions into groups that are assumed to have identical variances by applying \emph{$k$-means clustering} to the set $ \{ \log_{10}(\tilde{S}_{s, r}) : r = 1, \ldots, 27 \}$. The clustering is therefore based on the order of magnitude of the ensemble variability. The number of clusters, $k$, was chosen as the smallest $k$ such that the sum of the within-cluster sums-of-squares is at least 90\% of the total sums-of-squares. The fluxes derived from the LN observation type were used to do the clustering, and the same clusters were then used on the fluxes derived from the LG and the IS observation types. The resulting groupings of regions are shown in Figure~\ref{fig:clustering_map}; the group numbers are ordered from the clustered regions exhibiting the most variability (Group 1 has the highest average $\log_{10}(\tilde{S}_{s, r})$) to those exhibiting the least variability. Four clusters were chosen for DJF, JJA, and SON, and three clusters for MAM.

\end{document}